\documentclass[journal=jpcl,manuscript=article]{achemso}
\usepackage{chemformula} % Formula subscripts using \ch{}
\usepackage[T1]{fontenc} % Use modern font encodings
\usepackage{amsfonts}
\usepackage{bm}
\usepackage{multirow}
\usepackage{color}
\usepackage{multirow}
\usepackage{epsfig}
\usepackage{graphicx}
\usepackage[normalem]{ulem}
\usepackage{amsmath}
\usepackage{amssymb}
\usepackage{bm}
\usepackage{dcolumn}
\usepackage{braket}
\usepackage{bbold}
\usepackage{natbib}

\usepackage{array}
\usepackage{tabulary}
\usepackage{multirow}
\newcolumntype{C}[1]{>{\centering\arraybackslash}p{#1}}
\newcolumntype{L}[1]{>{\flushleft\arraybackslash}p{#1}}
\author{Xiaodong Zhou}
\affiliation
{Key Lab of Advanced Optoelectronic Quantum Architecture and Measurement (MOE), Beijing Key Lab of Nanophotonics Ultrafine Optoelectronic Systems, and School of Physics, Beijing Institute of Technology, Beijing 100081, China}
\altaffiliation{Contributed equally to this work}

\author{Run-Wu Zhang}
\affiliation
{Key Lab of Advanced Optoelectronic Quantum Architecture and Measurement (MOE), Beijing Key Lab of Nanophotonics Ultrafine Optoelectronic Systems, and School of Physics, Beijing Institute of Technology, Beijing 100081, China}
\altaffiliation{Contributed equally to this work}

\author{Zeying Zhang}
\affiliation
{Key Lab of Advanced Optoelectronic Quantum Architecture and Measurement (MOE), Beijing Key Lab of Nanophotonics Ultrafine Optoelectronic Systems, and School of Physics, Beijing Institute of Technology, Beijing 100081, China}

\author{Da-Shuai Ma}
\affiliation
{Key Lab of Advanced Optoelectronic Quantum Architecture and Measurement (MOE), Beijing Key Lab of Nanophotonics Ultrafine Optoelectronic Systems, and School of Physics, Beijing Institute of Technology, Beijing 100081, China}

\author{Wanxiang Feng}
\affiliation
{Key Lab of Advanced Optoelectronic Quantum Architecture and Measurement (MOE), Beijing Key Lab of Nanophotonics Ultrafine Optoelectronic Systems, and School of Physics, Beijing Institute of Technology, Beijing 100081, China}
\alsoaffiliation
{Peter Gr{\"u}nberg Institut and Institute for Advanced Simulation, Forschungszentrum J{\"u}lich and JARA, D-52425 J{\"u}lich, Germany}
\email{wxfeng@bit.edu.cn}

\author{Yuriy Mokrousov}
\affiliation{Peter Gr{\"u}nberg Institut and Institute for Advanced Simulation, Forschungszentrum J{\"u}lich and JARA, D-52425 J{\"u}lich, Germany}
\alsoaffiliation{Institute of Physics, Johannes Gutenberg University Mainz, 55099 Mainz, Germany}
\email{y.mokrousov@fz-juelich.de}

\author{Yugui Yao}
\affiliation
{Key Lab of Advanced Optoelectronic Quantum Architecture and Measurement (MOE), Beijing Key Lab of Nanophotonics Ultrafine Optoelectronic Systems, and School of Physics, Beijing Institute of Technology, Beijing 100081, China}
\email{ygyao@bit.edu.cn}

\title{Fully spin-polarized nodal loop semimetals in alkaline-metal monochalcogenide monolayers}

\begin{document}
	%%%%%%%%%%%%%%%%%%%%%%%%%%%%%%%%%%%%%%%%%%%%%%%%%%%%%%%%%%%%%%%%%%%%%
	%% The "entry" environment can be used to create an entry for the
	%\begin{tocentry}
	
	%\includegraphics[width=\columnwidth]{TOC}
	
	%Some journals require a graphical entry for the Table of Contents.
	%This should be laid out ``print ready'' so that the sizing of the
	%Inside the \texttt{tocentry} environment, the font used is Helvetica
	%8\,pt, as required by \emph{Journal of the American Chemical
	%The surrounding frame is 9\,cm by 3.5\,cm, which is the maximum
	%permitted for  \emph{Journal of the American Chemical Society}
	%graphical table of content entries. The box will not resize if the
	%content is too big: instead it will overflow the edge of the box.
	%
	%This box and the associated title will always be printed on a
	%separate page at the end of the document.
	%
	%\end{tocentry}
	
	
	\newpage
	
	\begin{abstract}
		Topological semimetals in ferromagnetic materials have attracted enormous attention due to the potential applications in spintronics.  Using the first-principles density functional theory together with an effective lattice model, here we present a new family of topological semimetals with a fully spin-polarized nodal loop in alkaline-metal monochalcogenide \textit{MX} (\textit{M} = Li, Na, K, Rb, Cs; \textit{X} = S, Se, Te) monolayers.  The half-metallic ferromagnetism can be established in \textit{MX} monolayers, in which one nodal loop formed by two crossing bands with the same spin components is found at the Fermi energy.  This nodal loop half-metal survives  even when considering the spin-orbit coupling owing to the symmetry protection provided by the $\mathcal{M}_{z}$ mirror plane.  The quantum anomalous Hall state and Weyl-like semimetal in this system can be also achieved by rotating the spin from the out-of-plane to the in-plane direction.  The \textit{MX} monolayers hosting rich topological phases thus offer an excellent materials platform for realizing the advanced spintronics concepts.
		
	\end{abstract}
	
	\newpage
	
	Spintronics is a multidisciplinary field, which utilizes the electron's spin degree of freedom as an information carrier for data storage and processing.~\cite{Wolf2001,Zutic2004}  Compared with conventional semiconductor devices, spintronic devices have higher integration density, faster processing speed, and lower power consumption.  However, many issues such as the generation and transport of a pure spin current still present profound challenges in realizing three-dimensional spintronics devices,~\cite{Awschalom2007,Felser2007} hindering the range of possible competitive applications.  In this respect, the  two-dimensional (2D) materials are being currently promoted as a flagship for realizing advanced spintronics concepts, among which the  emerging 2D topological quantum states exhibiting striking advantages for spintronics, have attracted tremendous attention in recent years.
	
	2D nodal loop semimetals (NLSs)~\cite{Burkov2011,R-Yu2017} provide an exciting avenue for realizing topological quantum phase transitions between the gapped and gapless states. Physically, the 2D NLSs are topologically protected by the crystal symmetry, e.g., mirror or glide-mirror symmetry. These proposed protection mechanisms provide guiding principles for predicting real materials to realize various NLSs. Many exotic properties were reported to be associated with NLSs, including high-temperature surface superconductivity,~\cite{Kopnin2011} non-dispersive Landau energy level,~\cite{Rhim2015} and specific long-range Coulomb interactions.~\cite{Huh2016}
	Recent advances in the domain of 2D ferromagnetic (FM) materials gave an additional boost to the young field of NLSs. Unlike the previously proposed NLSs in nonmagnetic materials, it was recently proposed that the NLSs can be achieved in 2D ferromagnets due to band crossings occuring between the states of opposite spins.~\cite{CW-Niu2018,BJ-Feng2019} Nevertheless, for applications which rely on strong spin-polarization of generated currents, it is  desired to realize the NLSs in fully spin-polarized 2D systems, such as half-metals.
	
	In this work, using the first-principles calculation together with an effective lattice model, we present a family of  fully spin-polarized NLSs, termed \textit{nodal loop half-metals (NLHMs)}, in 2D FM alkaline-metal monochalcogenides \textit{MX} (\textit{M} = Li, Na, K, Rb, Cs; \textit{X} = S, Se, Te) monolayers (MLs).  When the magnetization points along the  out-of-plane direction (i.e., $z$ axis), \textit{MX} MLs are 2D NLHMs in the sense that the nodal points intersecting from two spin-down bands form a closed loop in the first Brillouin zone.  The NLHM discovered here is a topologically protected state due to the presence of the $\mathcal{M}_z$ symmetry, and it is robust against the spin-orbit coupling (SOC).  The nodal loop will be gapped out if the magnetization is turned away from the $z$ axis, but interestingly, the emergent gapped state hosts the quantum anomalous Hall (QAH) effect characterized by the nonzero Chern number ($\mathcal{C}=\pm 1$).  When the magnetization lies within the crystal plane, a Weyl-like semimetal state, possessing two nodal points within a mirror plane, emerges.  Furthermore, the QAH and magneto-optical (MO) effects, providing the electric and optical means respectively, are proposed to detect such rich topological phases in the FM \textit{MX} MLs.  These findings  provide a new 2D material platform hosting the exotic NLHM and QAH states.

	\begin{figure}
		\includegraphics[width=0.8\columnwidth]{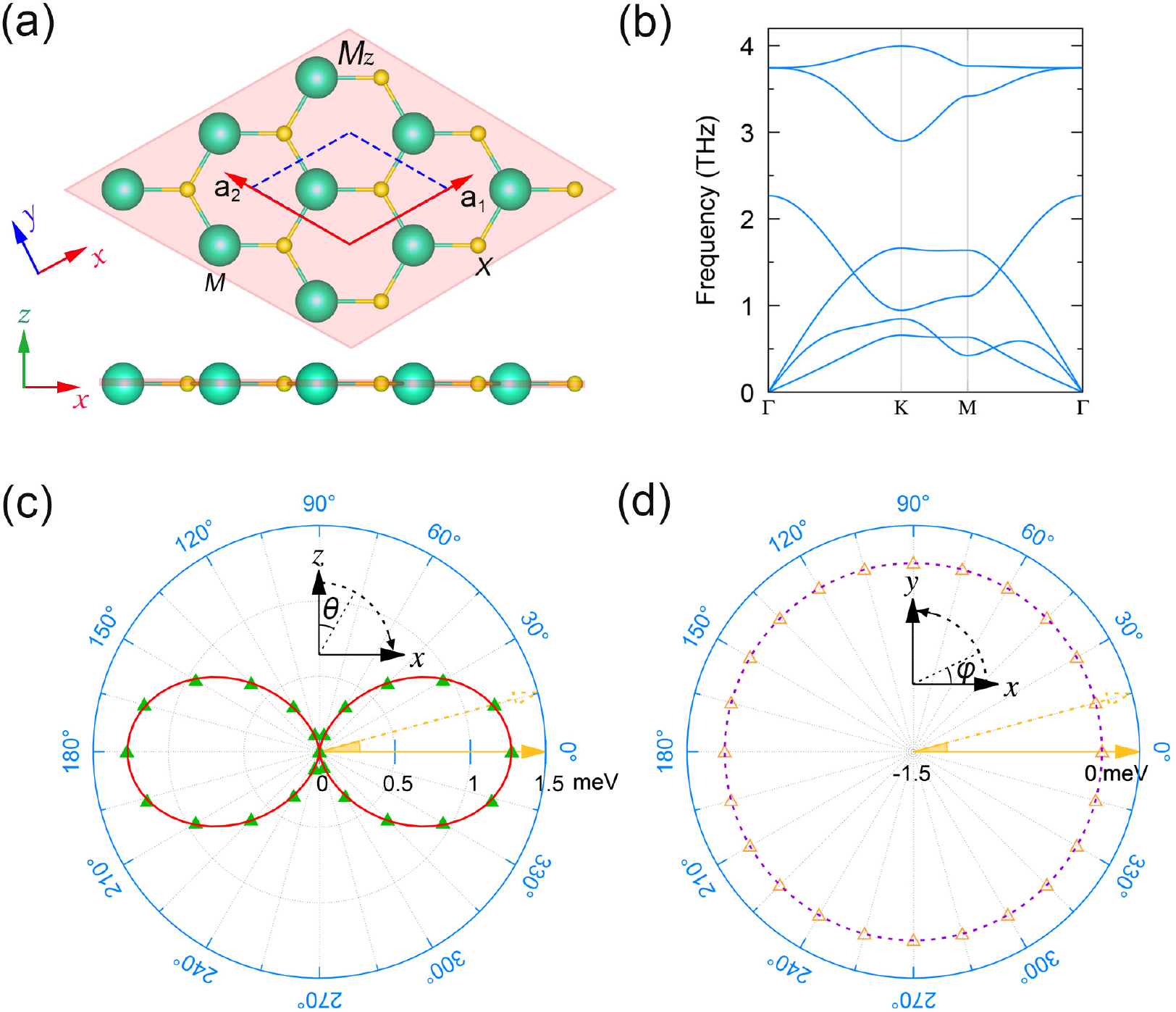}
		\caption{(a) The top and side views of \textit{MX} MLs.  The large green and small yellow balls represent \textit{M} and \textit{X} atoms, respectively.  The blue dashed lines indicate the 2D unit cell.  The mirror plane $\mathcal{M}_{z}$ (marked by pink color) is preserved if the spin magnetic moment points to the out-of-plane direction ($z$ axis).  (b) The phonon spectrum of CsS ML.  (c)(d) The calculated MAE by rotating the spin within $zx$ and $xy$ planes, respectively. The red solid line in (c) and violet dashed line in (d) represent the fitted curves, i.e., $\textrm{MAE} (\theta) = 1.32\times\textrm{cos}^{2}\theta-0.04\times\textrm{cos}^{4}\theta$ and  $\textrm{MAE} (\varphi) = 0$, respectively.}
		\label{fig:first}
	\end{figure}

	Recently, the alkaline-metal monochalcogenides \textit{MX} aroused considerable attention due to their unique FM half-metallicity.~\cite{GY-Gao2011,Moradi2012,Moradi2013,Rostami2013,Ahmadian2012,Senol2014,Afshari2015,Karaca2015,L-Li2015,G-Lei2016,Bialek2016,Rostami2016,Tabatabaeifar2017}  Bulk \textit{MX} exhibit various crystalline phases, such as CsCl, NiAs, Wurtzite, Zinc-blende, rocksalt structures.  In particular, CsS and CsSe with the CsCl-type structure [Figure S1(a), Supporting Information] are predicted to be most stable in energy.~\cite{Rostami2013,Karaca2015}  Taking CsS as an example, we checked its dynamical stability by calculating the phonon spectrum, which is free of imaginary frequencies [Figure S1(b)].  Interestingly, the FM half-metallicity is preserved at the (100), (110), and (111) surfaces of CsS.~\cite{Rostami2013} Owing to a strong interest in 2D ferromagnetism, one may naturally ask: can \textit{MX} MLs exist and will they exhibit observable half-metallicity?
	
	We focus on the (111) surface of bulk \textit{MX} and take CsS a prototypal example because other \textit{MX} shares very similar features.  As seen from Figure S1(a), the Cs and S atoms projecting onto the (111) plane are noncoplanar with a buckling height $h$=1.22 {\AA}.  When extracting the (111) plane from the bulk structure and relaxing the atomic positions in a slab model, the Cs and S atoms are inclined to form a coplanar structure, as shown in Figure~\ref{fig:first}(a).  One can see that CsS ML has a hexagonal honeycomb lattice which contains one Cs atom and one S atom.  The relaxed lattice constants of \textit{MX} MLs and the corresponding bulk counterparts are listed in Table S1 (Supporting Information).
	
	The practical feasibility for creating \textit{MX} MLs has been examined through the following three important aspects: (i) formation energy; (ii) dynamical stability; (iii) thermal stability. First of all, the formation energy of \textit{MX} MLs, defined as the energy (per atom) difference between ML structure and its bulk phase,~\cite{HL-Zhuang2012,HL-Zhuang2013,HL-Zhuang2014,HL-Zhuang2016,HL-Zhuang2016a} is calculated, and the results listed in Table S1 show that the experimental synthesis of \textit{MX} MLs is possible.  For example, the formation energy of CsS ML is 0.27 eV/atom, which is slightly larger than that of ZnO ML (0.19 eV/atom)~\cite{HL-Zhuang2013} but much smaller than that of silicene (0.76 eV/atom),~\cite{HL-Zhuang2012} whereas both ZnO ML and silicene have been recently experimentally synthesized.~\cite{Tusche2007,Vogt2012}  Additionally,  the dynamical stability of CsS ML is affirmed by the phonon spectrum, see Figure~\ref{fig:first}(b).  Moreover, the thermal stability of CsS ML is tested by using the molecular dynamics simulation, and the planar hexagonal honeycomb lattice maintains up to 100 K, see Figures S1(c) and S1(d).
	
	Magnetic properties are further explored in \textit{MX} MLs.  In the past, the long-range ferromagnetic order in low-dimensional systems (e.g., 2D case) has long been considered to be impossible due to the thermal fluctuations according to the Mermin-Wagner theorem.~\cite{Mermin_1966} However, this limitation has been challenged by the recent discoveries of intrinsic 2D ferromagnets, such as Cr$_{2}$Ge$_{2}$Te$_{6}$~\cite{C-Gong2017}, CrI$_{3}$~\cite{B-Huang2017}, and Fe$_{3}$GeTe$_{2}$~\cite{YJ-Deng2018}.  After doing a spin-polarized calculation, we learned that the spin magnetic moment carried by CsS ML is 1 $\mu_{B}$ per cell, being similar to its bulk phase.~\cite{Rostami2013} The ferromagnetism is mainly originated from the $p$ orbitals of S atom, whereas the spin magnetic moment on Cs atom can be negligible. In order to confirm this is an FM ground state, we compared the total energies between FM, antiferromagnetic and non-magnetic states in a doubled supercell.  The FM state is found to be more stable than antiferromagnetic and non-magnetic states with the smaller energies of 7.06 and 139.77 meV/cell, respectively.  Moreover, the SOC-induced magnetocrystalline anisotropy energy (MAE), defined by $\textrm{MAE}(\theta,\varphi)=E_{(\theta,\varphi)}-E_{(\theta=90^{\circ},\varphi=0^{\circ})}$ (where $\theta$ and $\varphi$ are polar and azimuthal angles), is analyzed by rotating the spin within both the $zx$ and $xy$ planes.  Figure~\ref{fig:first}(c) shows that the MAE of CsS ML in the $zx$ plane can be well fitted by an equation $\textrm{MAE}=K_{1}\textrm{cos}^{2}\theta+K_{2}\textrm{cos}^{4}\theta$, where $K_1$ (1.32 meV) and $K_2$ (-0.04 meV) are the magnetocrystalline anisotropy coefficients.~\cite{Getzlaff2007}  The positive value of MAE indicates a preferred magnetization along the $x$-axis rather than along the $z$-axis.  Figure~\ref{fig:first}(d) further illustrates that the MAE is isotropic in the $xy$ plane, suggesting that CsS ML belongs to the category of $XY$ ferromagnet.~\cite{HL-Zhuang2016}    The maximal value of MAE between out-of-plane and in-plane spin orientations reaches to 1.28 meV/cell, which is smaller than that of famous 2D ferromagnets CrI$_3$ (1.37 meV/cell)~\cite{WB-Zhang2015} and Fe$_3$GeTe$_2$ (2.76 meV/cell)~\cite{HL-Zhuang2016a}.  Interestingly, the magnetization direction in 2D ferromagnets can be effectively tuned by applying an external magnetic field, as experimentally realized in, e.g., Fe$_3$GeTe$_2$.~\cite{YJ-Deng2018}

	\begin{figure}
		\includegraphics[width=0.8\columnwidth]{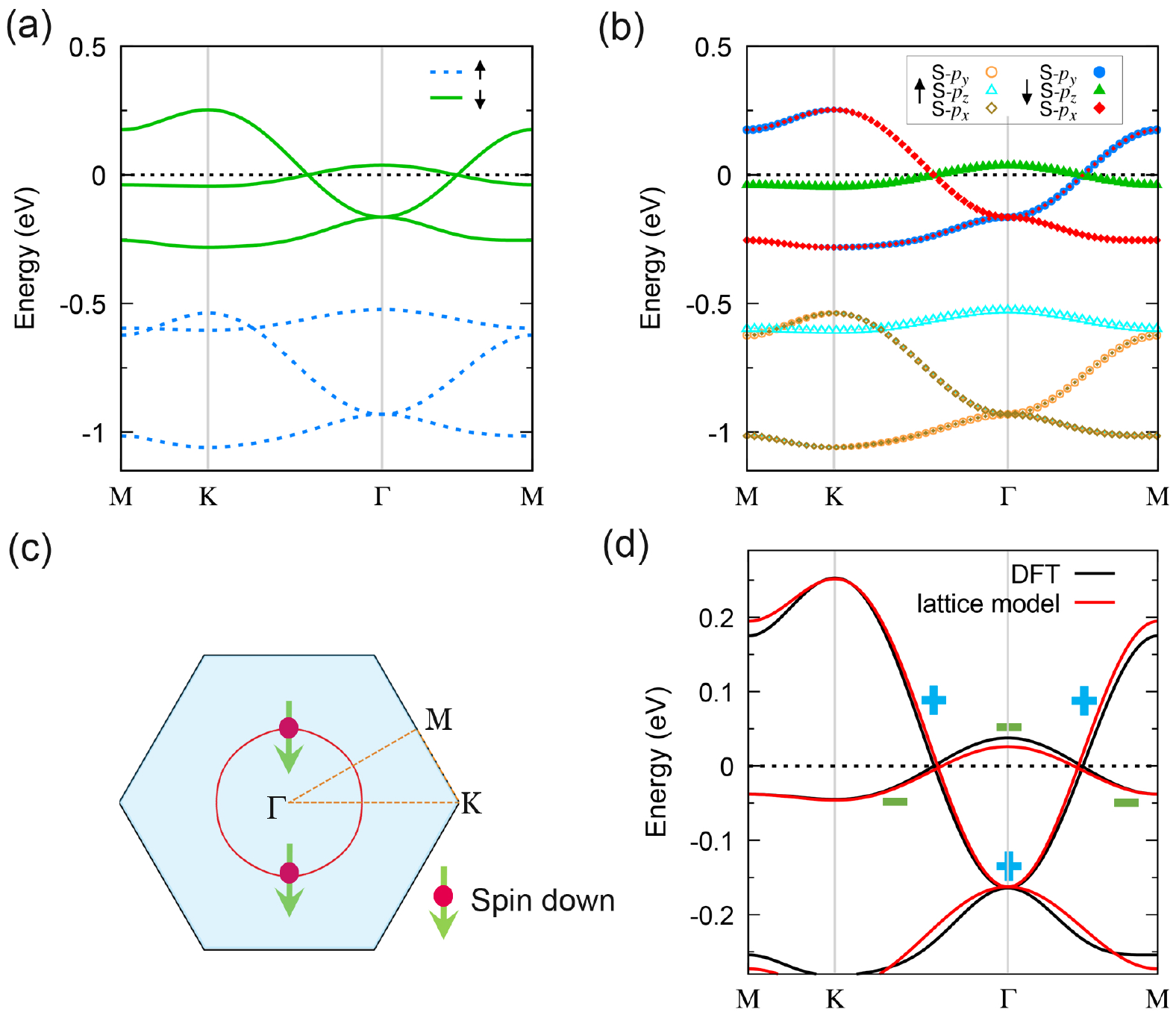}
		\caption{(a) The spin polarized band structure of CsS ML.  (b) The orbital-projected band structure of CsS ML.  (c) The momentum distribution of the band-crossing nodes near the Fermi energy.  (d) The band structures calculated by the DFT and the effective lattice model.  ``$+$'' and ``$-$'' mean the eigenvalues of $\mathcal{M}_{z}$ symmetry. SOC is not included in (a-d).}
		\label{fig:second}
	\end{figure}

	Computed without SOC, the emergent half-metallicity and the fully spin-polarized band crossings in \textit{MX} MLs are discussed.  In Figure~\ref{fig:second}(a), the spin-polarized band structure of CsS ML is plotted, in which the green solid (blue dashed) lines denote spin-down (spin-up) bands.  One can see that  CsS ML is a half-metal in the sense that the spin-down channel is metallic while the spin-up channel is semiconducting with a gap of about 3.79 eV.  Remarkably, the spin-down bands exhibit two band-crossing nodes at the Fermi energy along the $\Gamma$-K and $\Gamma$-M paths, respectively.  These bands are primarily derived from $p_{x, y}$ and $p_{z}$ orbitals of S atom, as shown in Figure~\ref{fig:second}(b).  In Figure~\ref{fig:second}(c), we display the momentum distribution of the band-crossing nodes, which form a Weyl-like (twofold-degenerate) nodal loop centered at the $\Gamma$ point in the first Brillouin zone.  Moreover, we calculated the $\mathcal{M}_{z}$ parities for the valence and conductance bands near the Fermi energy, and the opposite parities of two crossing bands [see Figure~\ref{fig:second}(d)] indicate that the gapless nodal loop is protected by the mirror reflection symmetry.
	
\begin{figure}
		\includegraphics[width=0.8\columnwidth]{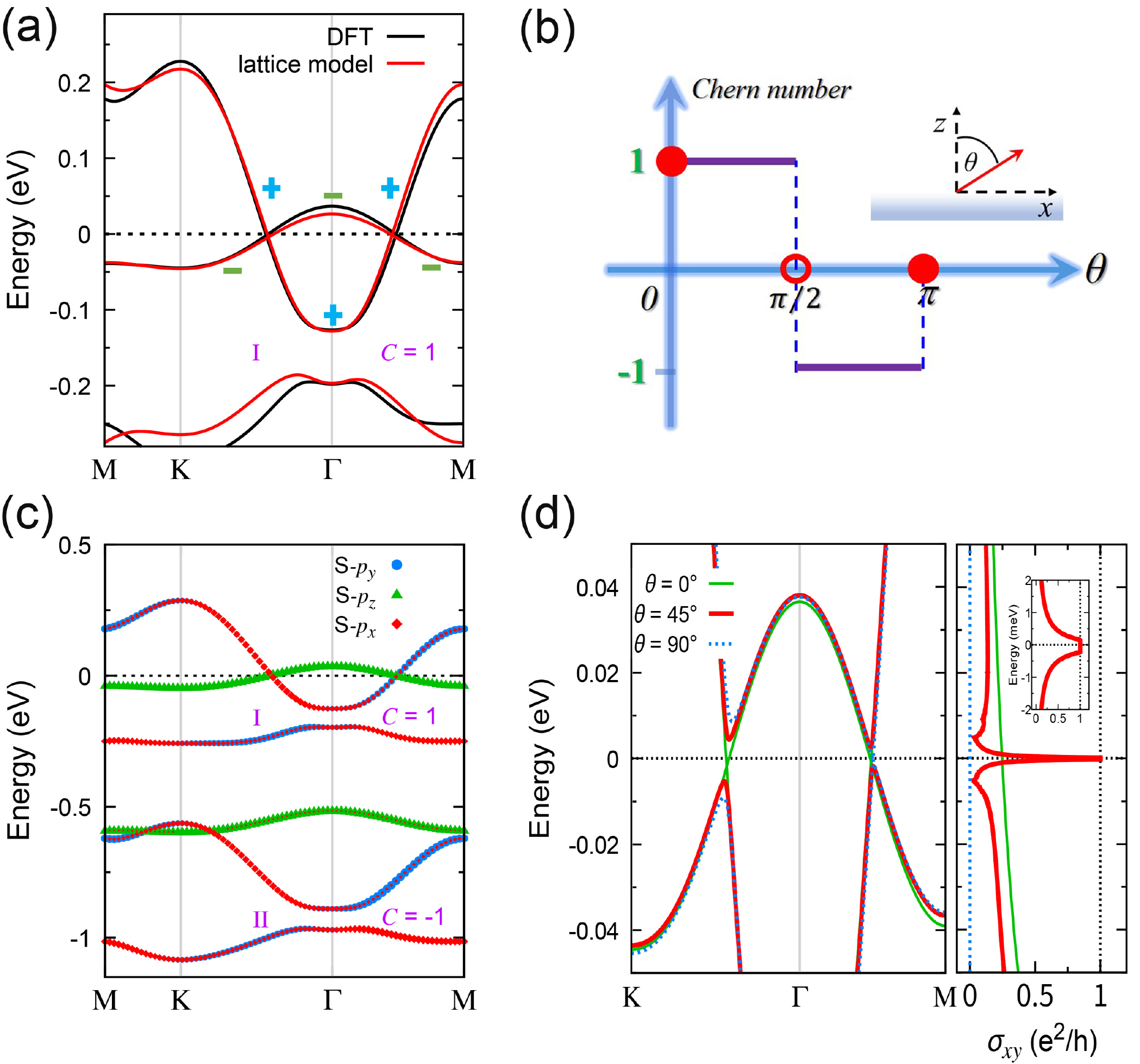}
		\caption{(a) The band structures of CsS ML calculated by the DFT and the effective lattice model with SOC.  (b) The calculated Chern number as a function of the magnetization direction $\theta$ (i.e., the polar angle away from the $z$-axis).  (c) The orbital-projected band structure of CsS ML with SOC.  The magnetization in (a) and (c) is considered along the out-of-plane direction.  After SOC is turned on, the nodal loop at the Fermi energy preserves, whereas two topologically nontrivial band gaps ($I$ and $II$) arise at the $\Gamma$ point, in which the AHC is quantized with the nonzero Chern numbers ($\mathcal{C}=+1$ and $\mathcal{C}=-1$).  (d) The band structures and anomalous Hall conductivities of CsS ML when $\theta=0^{\circ}$, $45^{\circ}$, and $90^{\circ}$.}
		\label{fig:third}
	\end{figure}

	Considering the SOC effect, the mirror reflection symmetry cannot protect the nodal loops in non-magnetic materials, referring to the previously predicted 2D NLSs~\cite{YJ-Jin2017,JL-Lu2017,CW-Niu2017,B-Yang2017,HH-Zhang2017,BJ-Feng2017,RW-Zhang2018,L-Gao2018,HY-Chen2018,P-Zhou2018}.  Different from the non-magnetic system, the nodal loop in FM CsS ML (with out-of-plane magnetization) survives after the SOC is turned on, see Figures~\ref{fig:third}(a) and~\ref{fig:third}(c).  The underlying mechanism is that in the case of out-of-plane magnetization $\mathcal{M}_{z}$ is a good symmetry regardless of SOC, and the unchanged $\mathcal{M}_{z}$ parities of the two crossing bands prevent the hybridization between $p_{x,y}$ and $p_{z}$ orbitals.  It shares a similar physical origin as the mixed nodal lines proposed in Ref.~\citenum{CW-Niu2018}.
	
	To capture the main physics in \textit{MX} ML with and without SOC, we developed an effective lattice model under a space group of $P\overline{6}m2$ (see Supporting Information).  The basis functions are chosen as $ \psi_1=\ket{p_{x}}$, $\psi_2=\ket{p_{y}}$, and $\psi_3=\ket{p_{z}}$ since the three bands near the Fermi energy are dominated by the $p$ orbitals of S atom.  The model parameters, including on-site energy, hopping integral, and SOC strength, are listed in Table S2.  The band structures calculated by the lattice model Hamiltonian reproduce well the low-energy electronic states obtained by the first-principles calculations, as shown in Figures~\ref{fig:second}(d) and~\ref{fig:third}(a).  In addition, we have identified all NLHM candidates in \textit{MX} MLs, including LiS, LiSe, LiTe, NaSe, and NaTe (see Figure S2).
	
	\begin{table}
		\caption{The magnetic point groups of \textit{MX} MLs as a function of the polar ($\theta$) or azimuthal ($\varphi$) angles when the spin is rotated within the $zx$ ($0\le\theta\le\pi$, $\varphi=0$) or $xy$ ($\theta=\pi/2$, $0\le\varphi<2\pi$) planes.}
		\label{tbl:mpg}
		\setlength{\tabcolsep}{3.6pt}
		\renewcommand{\arraystretch}{1.5}
		\begin{tabular}{lccccccccccccc}
			\hline
			\hline
			\multicolumn{1}{c}{$\theta$ or $\varphi$} & 
			\multicolumn{1}{c}{$0^{\circ}$} &
			\multicolumn{1}{c}{$15^{\circ}$} &
			\multicolumn{1}{c}{$30^{\circ}$} &
			\multicolumn{1}{c}{$45^{\circ}$} &
			\multicolumn{1}{c}{$60^{\circ}$} &
			\multicolumn{1}{c}{$75^{\circ}$} &
			\multicolumn{1}{c}{$90^{\circ}$} &
			\multicolumn{1}{c}{$105^{\circ}$} &
			\multicolumn{1}{c}{$120^{\circ}$} &
			\multicolumn{1}{c}{$135^{\circ}$} &
			\multicolumn{1}{c}{$150^{\circ}$} &
			\multicolumn{1}{c}{$165^{\circ}$} &
			\multicolumn{1}{c}{$180^{\circ}$} \\
			
			\hline
			$zx$  & $\bar{6}m'2'$  & $2'$ & $2'$ & $2'$ & $2'$ & $2'$ & $m'm2'$ & $2'$ & $2'$ & $2'$ & $2'$ & $2'$ & $\bar{6}m'2'$\\
			$xy$  & $m'm2'$ & $m'$ & $m'm'2$ & $m'$ & $m'm2'$ & $m'$ & $m'm'2$ & $m'$ & $m'm2'$ & $m'$ & $m'm'2$ & $m'$ & $m'm2'$\\
			\hline
			\hline
		\end{tabular}
	\end{table}

\begin{figure}
		\includegraphics[width=0.8\columnwidth]{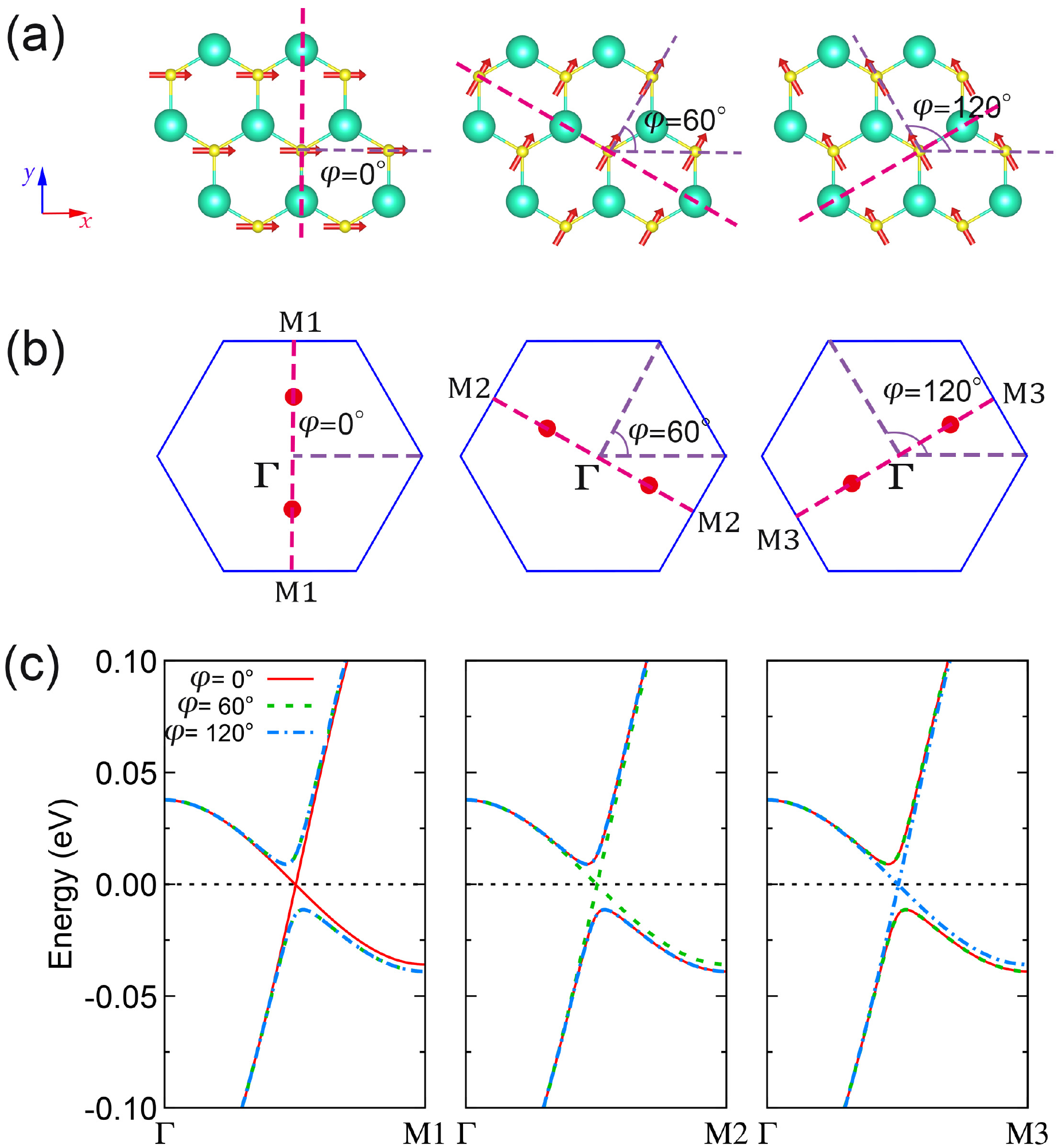}
		\caption{(a) The magnetic structures by varying the spin within the $xy$ plane.  $\varphi$ is the azimuthal angle starting from the $x$-axis.  The red dashed lines denote the mirror planes.  (b) The positions of one pair of two-fold degenerated points in the Brillouin zone.  (c) The relativistic band structures along the $\Gamma$-M1, $\Gamma$-M2, and $\Gamma$-M3 paths.}
		\label{fig:fourth}
	\end{figure}

	The abundant topological phases are usually expected by tuning the magnetization direction~\cite{Z-Liu2018,Hanke2017,ZY-Zhang2018}.  We first discuss the situation that rotates the spin within the $zx$ plane (i.e., $0\le\theta\le\pi$, $\varphi=0$).  The corresponding magnetic point groups computed by the \textsc{isotropy} code~\cite{Stokes} are listed in Table~\ref{tbl:mpg}.  One can see that the magnetic point group changes with a period of $\pi$ and only three non-repetitive elements need to be analyzed: $\bar{6}m^\prime2^\prime$ ($\theta = n\pi$), $m^\prime m2^\prime$ [$\theta = (n+1/2)\pi$], and $2^\prime$ [$\theta \neq n\pi$ and $\theta \neq (n+1/2)\pi$] with $n \in \mathbb{N}$.  As mentioned before, when the magnetization direction is out-of-plane [$\bar{6}m'2'$ ($\theta = n\pi$)], the nodal loop survives because the symmetry $\mathcal{M}_{z}$ is not broken.  It is well known that the Berry curvature $\boldsymbol{\Omega}=[\Omega^{x},\Omega^{y},\Omega^{z}]=[\Omega_{yz},\Omega_{zx},\Omega_{xy}]$ is a pseudovector.  Moreover, $\Omega^{x}$ and $\Omega^{y}$ are vanishing in 2D systems and only $\Omega^{z}$ ($=\Omega_{xy}$) is relevant for \textit{MX} MLs.  Since $\Omega_{xy}$ is an even function with respect to the mirror operation $\mathcal{M}_{z}$, the anomalous Hall conductivity (AHC), $\sigma_{xy}$, is expected to be nonzero in the case of $\theta = n\pi$, as shown in Figure~\ref{fig:third}(d).  When the magnetization direction is along the $x$-axis [$\theta=(n+1/2)\pi$], the group $m^\prime m2^\prime$ contains the mirror operation $\mathcal{M}_{x}$, which breaks the nodal loop except for two band crossing points sitting in the $\mathcal{M}_{x}$ plane, e.g., along the paths from the $\Gamma$ point to the $M$ and $-M$ points [see Figure~\ref{fig:third}(d) and Figure~\ref{fig:fourth}].  The symmetry $\mathcal{M}_{x}$ forces $\Omega_{xy}$ to be zero and consequently, there is $\sigma_{xy}=0$ [see Figure~\ref{fig:third}(d)]. In terminology of Ref.~\citenum{CW-Niu2018} these points present the mixed Weyl points of type-(i).  If $\theta \neq n\pi$ and $\theta \neq (n+1/2)\pi$, the group $2'$ does not have any mirror operation, and therefore all band crossing points are gapped out.  Importantly, the group $2'$ contains a $\mathcal{TC}_{2y}$ operation ($\mathcal{T}$ is the time-reversal symmetry and $\mathcal{C}_{2y}$ is the two-fold rotation around the $y$-axis), which in principle renders non-vanishing $\Omega_{xy}$.  Seen from Figure~\ref{fig:third}(d), the AHC $\sigma_{xy}$ is indeed nonzero and strikingly, the quantized value, $\sigma_{xy}=e^2/h$ (here, taking $\theta=45^{\circ}$ as an example), suggests a QAH insulator with the Chern number $\mathcal{C}=+1$.  The evaluation of topological phase as a function of the polar angle $\theta$ is sketched in Figure~\ref{fig:third}(b).
	
	Next, we explore the possible topological phases when the spin lies within the $xy$ plane ($\theta=\pi/2$, $0\le\varphi<2\pi$).  From Table~\ref{tbl:mpg}, one can observe that the magnetic point group has a period of $\pi/3$ with respect to $\varphi$: $m^{\prime}m2^{\prime} \Rightarrow m^{\prime} \Rightarrow m^{\prime}m^{\prime}2 \Rightarrow m^{\prime} \Rightarrow m^{\prime}m2^{\prime}$. Particularly, the group $m^{\prime}m2^{\prime}$ ($\theta = n\pi/3$) has a mirror plane that is perpendicular to the spin direction, as shown in Figure~\ref{fig:fourth}(a).  This mirror plane drives the nodal loop into one pair of two-fold degenerated points sitting in the mirror plane, see Figures~\ref{fig:fourth}(b) and~\ref{fig:fourth}(c).  If such a mirror symmetry is absent [$m^{\prime}$ and $m^{\prime}m^{\prime}2$, ($\theta \neq n\pi/3$)], the nodal loop will be fully gapped out and the insulating phase is topologically trivial due to the presence of $\mathcal{TM}_{z}$ symmetry. The reason is that $\Omega_{xy}$ is odd (even) under the symmetry $\mathcal{T}$ ($\mathcal{M}_{z}$) and thus is odd under the symmetry $\mathcal{TM}_{z}$.  By integrating $\Omega_{xy}$ in the entire Brillouin zone, $\sigma_{xy}$ has to be zero, giving the Chern number $\mathcal{C}=0$.  This is different from the scenario of the QAH insulator with in-plane magnetization proposed by Liu \textit{et al.},~\cite{Z-Liu2018} because in their systems (e.g., LaCl monolayer) the $\mathcal{TM}_{z}$ symmetry lacks.

\begin{figure}
		\includegraphics[width=0.8\columnwidth]{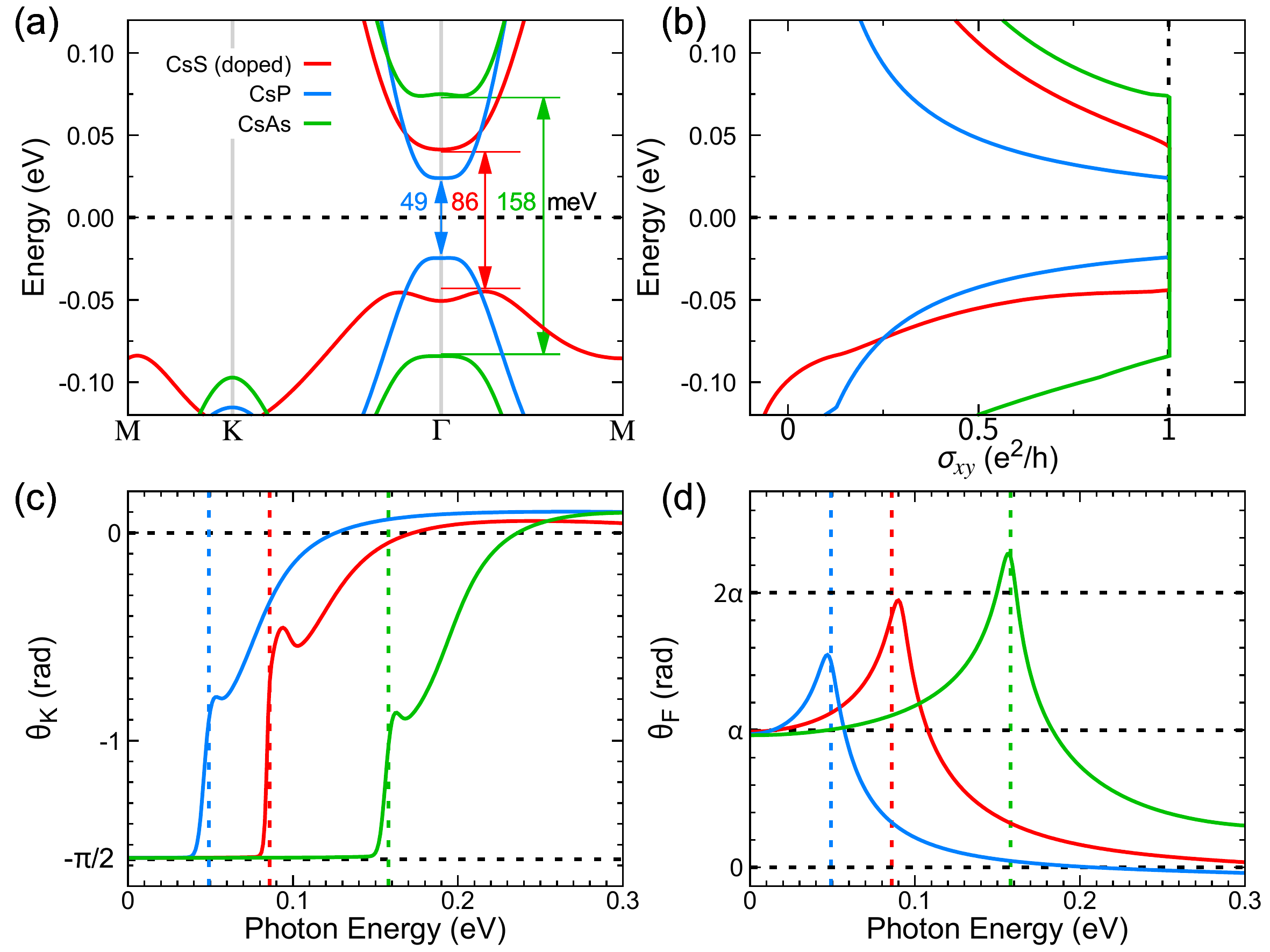}
		\caption{(a) Relativistic band structures, (b) anomalous Hall conductivities, (c) Kerr rotation angles, and (d) Faraday rotation angles of CsS (doped one hole), CsP, and CsAs MLs.  The dashed vertical lines in (c) and (d) denote the topologically nontrivial band gaps.  The $\alpha$ in (d) is fine structure constant.}
		\label{fig:fifth}
	\end{figure}

	QAH effect plays an important role in the application of spintronics. Herein, the QAH state can also be found in CsS ML even the magnetization takes an out-of-plane direction.  After SOC is switched on, two topologically nontrivial band gaps (\textit{I} and \textit{II}) arise at the $\Gamma$ point, in which the AHC is quantized with the Chern numbers $\mathcal{C}=+1$ and $\mathcal{C}=-1$ [see Figures~\ref{fig:third}(a) and~\ref{fig:third}(c)].  The gap \textit{I} is close to the Fermi energy and the electrostatic doping could be used to engineer the electronic state.  Indeed, if one hole is doped into CsS ML, a QAH state appears with a large nontrivial band gap of 86 meV [see Figures~\ref{fig:fifth}(a) and~\ref{fig:fifth}(b)].  Note that doping one hole into CsS ML corresponds to a doping concentration of $3.52\times 10^{14}$ cm$^{-2}$, which is achievable by the current experimental techniques.~\cite{Efetov2010,JT-Ye2011}  Similar to CsS, CsP and CsAs host rich bulk phases and exhibit FM half-metallicity on the low-indexed surfaces~\cite{Y-Zhang2012,Lakdja2013,XP-Wei2014,Lakdja2014,XP-Wei2015,Q-Gao2015}.  The dynamical stability of CsP and CsAs MLs has been testified by calculating their phonon spectra, shown in Figure S3, which indicate the practical feasibility. Since CsP and CsAs are isostructural to CsS but one hole is less in their native states, we propose that CsP and CsAs MLs are intrinsic QAH insulators [see Figures~\ref{fig:fifth}(a) and~\ref{fig:fifth}(b)].
	
	In addition to the anomalous Hall effect, the MO Kerr and Faraday effects, being a kind of non-contact (non-damaging) optical technique, are powerful tools for measuring the magnetism in 2D materials.~\cite{C-Gong2017,B-Huang2017}  Regarding the QAH insulator systems, an additional magnetoelectric term $(\Theta\alpha/4\pi^{2})$\textbf{E}$\cdot$\textbf{B} (here $\Theta$ is magnetoelectric polarizability and $\alpha = e^{2}/\hbar c$ is fine structure constant) should be added into the usual Lagrangian such that the Maxwell's equations have to be modified.~\cite{XL-Qi2008}  In the low-frequency limit (i.e., $\omega \ll  E_{g}/\hbar$, where $E_{g}$ is the topologically nontrivial band gap), the MO Kerr and Faraday rotation angles are quantized to $\theta_{K}\simeq -\pi/2$ and $\theta_{F}\simeq \mathcal{C}\alpha$ ($\mathcal{C}$ is the Chern number), respectively.~\cite{Maciejko2010,Tse2010}  Therefore, by an optical way, the modern MO techniques have been used to characterize the QAH state.~\cite{L-Wu2016,Okada2016,Shuvaev2016,Dziom2017}  Figures~\ref{fig:fifth}(c) and~\ref{fig:fifth}(d) plot the Kerr and Faraday spectra of CsS, CsP, and CsAs MLs, from which one can clearly find the quantized behaviors of Kerr and Faraday rotation angles in the low-frequency limit.
	
	In conclusion, using the first-principles calculations and an effective lattice model, we propose a class of nodal loop half-metal candidates in ferromagnetic alkaline-metal monochalcogenides \textit{MX} (\textit{M} = Li, Na, K, Rb, Cs; \textit{X} = S, Se, Te) monolayers.  The nodal loop is formed by two crossing bands with the same spin components, and in the case of out-of-plane magnetization the nodal loop is protected by the symmetry $\mathcal{M}_{z}$ that is not influenced by spin-orbit coupling.  The topological phase transition from a nodal loop half-metal, to a quantum anomalous Hall insulator, and to a Weyl-like semimetal is found by rotating the spin from the $z$-axis to the $x$-axis.  For the in-plane magnetization, the Weyl-like semimetal appears in a period of $\pi/3$ starting from the $x$-axis, and the intermediate phases are topologically trivial insulators.  Moreover, we show that the quantum anomalous Hall state can also be realized in CsS ML by doping one hole or by replacing S atom with P or As atoms.  The exotic quantum MO Kerr and Faraday effects are expected in CsS, CsP, and CsP MLs.  Our work reveals a kind of novel 2D ferromagnets hosting rich topological phases, which serves as good candidates for the promising spintronics.
	
	\textit{Note added:} After completed the bulk of our work, we became aware of two related works that predict 2D nodal loop half-metals in MnN~\cite{SS-Wang2019} and PtCl$_3$~\cite{JY-You2019} monolayers.

	\begin{suppinfo}
		
		The Supporting Information contains: (a) the details of first-principles calculations, (b) an effective lattice model, and (c) supplementary tables and figures.
		
	\end{suppinfo}
	
	\begin{acknowledgement}
		The authors thank the fruitful discussions with Branton J. Campbell.  W.F and Y.Y acknowledge the support from the National Natural Science Foundation of China (Grant Nos. 11874085, 11734003, and 11574029) and the National Key R\&D Program of China (Grant No. 2016YFA0300600).  W. F. also acknowledges the funding through an Alexander von Humboldt Fellowship. Y.M. acknowledges the funding under SPP 2137 ``Skyrmionics" (project MO 1731/7-1), collaborative Research Center SFB 1238, and project  MO  1731/5-1 of Deutsche Forschungsgemeinschaft (DFG).  We acknowledge computing time on the supercomputers JUQUEEN and JURECA at J\"ulich Supercomputing Centre and JARA-HPC of RWTH Aachen University.
	\end{acknowledgement}
	
	%\bibliography{refs}

	\providecommand{\latin}[1]{#1}
	\makeatletter
	\providecommand{\doi}
	{\begingroup\let\do\@makeother\dospecials
		\catcode`\{=1 \catcode`\}=2 \doi@aux}
	\providecommand{\doi@aux}[1]{\endgroup\texttt{#1}}
	\makeatother
	\providecommand*\mcitethebibliography{\thebibliography}
	\csname @ifundefined\endcsname{endmcitethebibliography}
	{\let\endmcitethebibliography\endthebibliography}{}

	\newpage

	\setcounter{figure}{0}
	\setcounter{table}{0}
	\makeatletter 
	\renewcommand{\thefigure}{S\@arabic\c@figure}
	\renewcommand{\thetable}{S\arabic{table}}
	\makeatother
	
	\section{Supporting Information for ``Fully spin-polarized nodal loop semimetals in alkaline-metal monochalcogenide monolayers''}
	
	\subsection{Details of first-principles calculation}
	
	The first-principles density functional theory calculations are performed using the projected augmented wave method~\cite{Bloechl1994}, as implemented in the Vienna \textit{ab initio} simulation package (\textsc{vasp})~\cite{Kresse1996a}. The Perdew-Burke-Ernzerhof parameterized generalized-gradient approximation (PBE-GGA)~\cite{Perdew1996} is adopted to treat the exchange-correlation effect.  The energy cut-off of the plane wave with 500 eV and the $k$-mesh of 25$\times$25$\times$1 are used in the self-consistent field calculations.  The convergence criteria of force and energy are chosen to be 0.005 eV/{\AA} and 10$^{-8}$ eV, respectively.  A vacuum region with a thickness of 16 {\AA} is used to avoid the interactions between adjacent slabs. The phonon spectrum is performed based on the density functional perturbation theory (DFPT)~\cite{Baroni2001}.  A penalty functional is added in order to fix the direction of the local spin moment. After obtaining the converged ground state charge density, the mostly localized Wannier functions by projecting the $p$-orbitals of S atom are constructed on a $k$-mesh of 10$\times$10$\times$1, using the \textsc{wannier90} package~\cite{Mostofi2008}.  Then, the intrinsic anomalous Hall conductivity is evaluated on an extremely dense $k$-mesh of 3000$\times$3000$\times$1, using the Kubo formula,~\cite{WX-Feng2016}
	\begin{eqnarray}
		\sigma_{xy} &=& \mathcal{C}\frac{e^{2}}{h}, \label{eq:sigma} \\
		\mathcal{C} &=& \frac{1}{2\pi}\sum_{n}\int_{\textrm{2D-BZ}}\Omega^{n}_{xy}(\bm{k})d^{2}k, \label{eq:chern} \\
		\Omega^{n}_{xy}(\bm{k}) &=& -\sum_{n^{\prime} \neq n} \frac{2\textrm{Im}\left\langle \psi_{n\bm{k}}\right|\hat{v}_{x}\left| \psi_{n^{\prime}\bm{k}} \right\rangle \left\langle \psi_{n^{\prime}\bm{k}} \right|\hat{v}_{y}\left|\psi_{n\bm{k}} \right\rangle}{\left(\omega_{n^{\prime}\bm{k}}-\omega_{n\bm{k}}\right)^{2}}. \label{eq:berry}
	\end{eqnarray}
	Here, $\mathcal{C}$ is the Chern number, $\Omega^{n}_{xy}(\bm{k})$ is the Berry curvature, $v_{x}$ and $v_{y}$ are velocity operators, $\psi_{n\textbf{\textit{k}}}$ and $\hbar\omega_{n\bm{k}}=\epsilon_{n\bm{k}}$ are eigenvector and eigenvalue at the band index $n$ and momentum $\bm{k}$, respectively.
	
	Extending the anomalous Hall effect to the ac case, the optical Hall conductivity,~\cite{Mostofi2008}
	\begin{equation}
		\sigma_{xy}(\omega) = \frac{ie^{2}\hbar}{N_{k}V_{c}} \sum_{\bm{k}} \sum_{n,m} \frac{f_{m\bm{k}}-f_{n\bm{k}}}{\epsilon_{m\bm{k}}-\epsilon_{n\bm{k}}} \frac{\langle \psi_{n\bm{k}} | \upsilon_{x} | \psi_{m\bm{k}} \rangle \langle \psi_{m\bm{k}} | \upsilon_{y} | \psi_{n\bm{k}} \rangle}{\epsilon_{m\bm{k}}-\epsilon_{n\bm{k}}-(\hbar\omega+i\eta)}, \label{eq:optic}
	\end{equation}
	is calculated on the same $k$-mesh of 3000$\times$3000$\times$1.  Here, $V_{c}$ is the cell volume, $N_{k}$ is the number of $k$-points used for sampling the first Brillouin zone, $\hbar\omega$ is the phonon energy, and $\eta$ is the smearing parameter.
	
	The magneto-optical Kerr and Faraday angles in the quantum anomalous Hall insulator are given by,~\cite{Tse2011}
	\begin{eqnarray}
		\theta_{K} &=& \frac{1}{2}(\textrm{arg}\lbrace E^{r}_{+} \rbrace-\textrm{arg}\lbrace E^{r}_{-} \rbrace), \label{eq:kerr} \\
		\theta_{F} &=& \frac{1}{2}(\textrm{arg}\lbrace E^{t}_{+} \rbrace-\textrm{arg}\lbrace E^{t}_{-} \rbrace), \label{eq:fara} 
	\end{eqnarray}
	where $E^{r, t}_{\pm} = E^{r, t}_{x} \pm iE^{r, t}_{y}$ are the left ($+$) and right ($-$) circularly polarized components of the outgoing reflected ($r$) and transmitted ($t$) electric fields.  When the thickness of the quantum anomalous Hall insulators is much shorter than the wavelength of incident light, the outgoing electric fields can be written as,~\cite{Tse2011}
	\begin{eqnarray}
		E^{r}_{x} &=& [1-(1+4\pi\sigma_{xx})^{2}-(4\pi\sigma_{xy})^{2}]A, \label{eq:Ex_r} \\ 
		E^{r}_{y} &=& 8\pi\sigma_{xy}A, \label{eq:Ey_r} \\
		E^{t}_{x} &=& 4(1+2\pi\sigma_{xx})A, \label{eq:Ex_t} \\
		E^{t}_{y} &=& 8\pi\sigma_{xy}A, \label{eq:Et_y}
	\end{eqnarray}
	with $A = 1/[(2+4\pi\sigma_{xx})^{2}+(4\pi\sigma_{xy})^{2}]$.  Here, $\sigma_{xx}$ and $\sigma_{xy}$ are the diagonal and off-diagonal elements of optical conductivity.  In the low-frequency limit ($\omega\rightarrow0$), the optical conductivity in quantum anomalous Hall insulators behaves like $\sigma^{R}_{xx}=0$, $\sigma^{I}_{xx}=0$, $\sigma^{R}_{xy}=\mathcal{C}e^{2}/h$, and $\sigma^{I}_{xy}=0$, ($\mathcal{C}$ is the Chern number, the superscripts $R$ and $I$ represent the real and imaginary parts, respectively).  Then, the Equations~\eqref{eq:kerr}-\eqref{eq:Et_y} can be simplified to,~\cite{Tse2011,Tse2010,Maciejko2010}
	\begin{eqnarray}
		\theta_{K} &=& -\textrm{tan}^{-1}[c/(2\pi\sigma^{R}_{xy})] = -\textrm{tan}^{-1}(1/C\alpha) \simeq -\pi/2,  \label{eq:kerr2} \\
		\theta_{F} &=& \textrm{tan}^{-1}(2\pi\sigma^{R}_{xy}/c) = \textrm{tan}(\mathcal{C}\alpha) \simeq \mathcal{C}\alpha. \label{eq:fara2}
	\end{eqnarray}
	Therefore, in the low-frequency limit, the magneto-optical Kerr and Faraday angles in quantum anomalous Hall insulators are quantized to $\pi/2$ and $\mathcal{C}\alpha$, respectively.

	\subsection{Lattice model Hamiltonian}
	
	To capture the main physics in \textit{MX} MLs with and without spin-orbit coupling, we developed an effective lattice model under the space group $P\overline{6}m2$ (No.187).  Using the first-principles calculations, we know that the $p$ orbitals of S atom are the dominant components for the conduction and valence bands near the Fermi energy.  Neglecting SOC first and considering the three states $\psi_1=\ket{p_{x}}$, $\psi_2=\ket{p_{y}}$, and $\psi_3=\ket{p_{z}}$ as the bases, the lattice model Hamiltonian is expressed as:
	\begin{eqnarray}
		H_{ij}(\boldsymbol{k})&=&\sum_{\boldsymbol{R}}e^{i\boldsymbol{k \cdot R}}t_{ij}(\boldsymbol{R}), \\
		t_{ij}(\boldsymbol{R})&=&\braket{\psi_{i}(\boldsymbol{R})|H|\psi_{j}(\boldsymbol{r}-\boldsymbol{R})},
	\end{eqnarray}
	where $t_{ij}$ denotes the hopping integral between the neighboring sites with a displacement $\mathbf{R}$, satisfying $t_{ij}(R\boldsymbol{R}_n)=D_i(R)t_{ij}(\boldsymbol{R}_n)[D_j(R)]^{\dag}$. Here, the $D_i(R)$ is the matrix of the $i$th irreducible representation.  The Hamiltonian can be explicitly written as:
	\begin{equation}
		\label{hamiltonian}
		H(\boldsymbol{k})=\left[
		\begin{matrix}
			H_{11}&H_{12}&0\\
			H_{21}&H_{22}&0\\
			0&0&H_{33}\\
		\end{matrix}
		\right],
	\end{equation}
	with the matrix elements,
	\begin{equation}
		\begin{split}
			H_{11}=&e_1+\frac{1}{2} t_{11} \cos (k_x+k_y)+\frac{3}{2} t_{22} \cos (k_x+k_y)+2 t_{11} \cos k_x+\frac{1}{2} t_{11} \cos k_y+\frac{3}{2} t_{22} \cos k_y, \\
			H_{12}=&\frac{1}{2} \sqrt{3} t_{11} \cos (k_x+k_y)+2 i t_{21} \sin (k_x+k_y)-\frac{1}{2} \sqrt{3} t_{22} \cos (k_x+k_y), \\
			&-2 i t_{21} \sin k_x-\frac{1}{2} \sqrt{3}
			t_{11} \cos k_y-2 i t_{21} \sin k_y+\frac{1}{2} \sqrt{3} t_{22} \cos k_y, \\
			H_{22}=&e_1+\frac{3}{2} t_{11} \cos (k_x+k_y)+\frac{1}{2} t_{22} \cos (k_x+k_y)+2 t_{22} \cos k_x+\frac{3}{2} t_{11} \cos k_y+\frac{1}{2} t_{22} \cos k_y, \\
			H_{33}=&e_2+2 r_{11} (\cos (k_x+k_y)+\cos k_x+\cos k_y), \\
			H_{ij}=&H^*_{ji}.
		\end{split}
		\label{eq:ham}
	\end{equation}
	The $\sigma_h$ mirror plane in $D_{3h}$ point group is responsible for the avoiding hybridization between $p_{x,y}$ and $p_{z}$ orbitals due to their opposite eigenvalues with respect to $\sigma_h$.  The model parameters obtained by fitting the first-principles band
	structures are listed in Table~\ref{tbl:model}.
	
	Since the topological phase transition can be triggered by changing magnetization direction, a general direction of spin quantization $\bf{m}$, characterized by polar ($\theta$) and azimuthal ($\varphi$) angles, is necessary to be considered.  Here, the spinors are defined by:
	\begin{eqnarray}
		\ket{\uparrow}_{\bf{m}} &=& e^{-i \frac{\varphi}{2}} \cos \frac{\theta}{2}\ket{\uparrow}+e^{i \frac{\varphi}{2}} \sin \frac{\theta}{2}\ket{\downarrow}, \\
		\ket{\downarrow}_{\bf{m}} &=& -e^{-i \frac{\varphi}{2}} \sin \frac{\theta}{2}\ket{\uparrow}+e^{i \frac{\varphi}{2}} \cos \frac{\theta}{2}\ket{\downarrow},
	\end{eqnarray}
	where $0\leq\theta\leq\pi$, $0\leq\varphi<2\pi$, and $\ket{\uparrow}$ and $\ket{\downarrow}$ are the eigenvectors of spin operator $\hat{s}_{z}$.  The SOC term of Hamiltonian is written as:
	\begin{equation}
		H^{\textrm{SOC}}_{ij\sigma\sigma\prime}=\frac{\xi}{2}\braket{\psi_{i\sigma}|\boldsymbol{L\cdot S}|\psi_{j\sigma'}}.
	\end{equation}
	For the case of $\varphi=0$ (i.e., rotating the spin within the $zx$ plane), it turns out be a function of polar angle $\theta$,
	\begin{equation}
		H^{\textrm{SOC}}=\frac{\xi}{2}\left[
		\begin{matrix}
			0 & -i \cos \theta & 0 & 0 & i \sin \theta & 1 \\
			i \cos \theta & 0 & -i \sin \theta & -i \sin \theta & 0 & -i \cos \theta \\
			0 & i \sin \theta & 0 & -1 & i \cos \theta & 0 \\
			0 & i \sin \theta & -1 & 0 & i \cos \theta & 0 \\
			-i \sin \theta & 0 & -i \cos \theta & -i \cos \theta & 0 & i \sin \theta \\
			1 & i \cos \theta & 0 & 0 & -i \sin \theta & 0 \\
		\end{matrix}
		\right].
	\end{equation}
	There exists three topological phases by varying the polar angle $\theta$ in the $zx$ plane:  (i) when $\theta = n\pi$ ($n\in\mathbb{N}$), a nodal loop half-metal protected by $\mathcal{M}_{z}$ symmetry emerges; (ii) when $\theta = (n+1/2)\pi$ ($n\in\mathbb{N}$), a Weyl-like semimetal, which hosts two nodal points in the plane normal to the $x$-axis, appears because the magnetic structure in this case has $\mathcal{M}_{x}$ symmetry;  (iii) when $\theta \neq n\pi$ and $\theta \neq (n+1/2)\pi$ ($n\in\mathbb{N}$), since all mirror symmetries are broken, the band crossing points are fully gapped out and the system evolves into a quantum anomalous Hall state with the Chern number $\mathcal{C}=\pm1$.

	\newpage
	
	\subsection{Supplementary tables and figures}

	\begin{table}
		\caption{The lattice constants $a_{m}$ ($a_{b}$) and bond lengths $d_{m}$ ($d_{b}$) for the monolayer (bulk) of alkaline-metal monochalcogenides \textit{MX} (\textit{M} = Li, Na, K, Rb, Cs; \textit{X} = S, Se, Te) as well as the isostructural CsP and CsAs.  The formation energy of monolayer is defined as $E_{f}=E_{\textrm{2D}}/N_{\textrm{2D}}-E_{\textrm{3D}}/N_{\textrm{3D}}$, where $E_{\textrm{2D}}$ ($E_{\textrm{3D}}$) and $N_{\textrm{2D}}$ ($N_{\textrm{3D}}$) are total energy and the number of atoms in monolayer (bulk) structure, respectively.  The magnetocrystalline anisotropy energy (MAE) of monolayer is defined as the energy difference between the out-of-plane and in-plane magnetization directions.}
		\label{tbl:fp}
		\begin{tabular}{ccccccc}
			\hline
			\hline
			\multicolumn{1}{c}{ \multirow{2}{*}{Structure}} & 
			\multicolumn{1}{c}{$a_{m}$} &
			\multicolumn{1}{c}{$a_{b}$} &
			\multicolumn{1}{c}{$d_{m}$} &   
			\multicolumn{1}{c}{$d_{b}$} &   
			\multicolumn{1}{c}{$E_{f}$}  &   
			\multicolumn{1}{c}{MAE}  \\
			
			\multicolumn{1}{c}{} & 
			\multicolumn{1}{c}{(\AA)} &
			\multicolumn{1}{c}{(\AA)} &
			\multicolumn{1}{c}{(\AA)} &   
			\multicolumn{1}{c}{(\AA)} &   
			\multicolumn{1}{c}{(eV/atom)}  &   
			\multicolumn{1}{c}{(meV/cell)} \\
			
			\hline
			LiS  &  4.08 & 3.11  & 2.36 & 2.69 & 0.43 & 0.48 \\
			LiSe  &  4.30 & 3.29  & 2.48 & 2.85 & 0.46 & 5.90 \\
			LiTe  &  4.65 & 3.60  & 2.69 & 3.12 & 0.43 & 8.38 \\
			\hline
			NaS  &  4.70 & 3.41  & 2.72 & 2.95 & 0.40 & 0.63 \\
			NaSe  &  4.92 & 3.57  & 2.84 & 3.09 & 0.46 & 2.59 \\
			NaTe  &  5.28 & 3.85  & 3.05 & 3.33 & 0.46 & 1.55 \\
			\hline
			KS  &  5.34 & 3.81  & 3.08 & 3.30 & 0.33 & 0.59 \\
			KSe  &  5.57 & 3.95  & 3.21 & 3.42 & 0.36 & 0.13 \\
			KTe  &  5.96 & 4.19  & 3.44 & 3.63 & 0.41 & -2.14 \\
			\hline
			RbS  &  5.56 & 4.01  & 3.21 & 3.47 & 0.29 & 0.63 \\
			RbSe  &  5.80 & 4.14  & 3.35 & 3.58 & 0.33 & 0.18 \\
			RbTe  &  6.21 & 4.37  & 3.58 & 3.79 & 0.36 & -2.14 \\
			\hline
			CsS  &  5.73 & 4.22  & 3.31 & 3.66 & 0.27 & 1.28 \\
			CsSe  &  6.00 & 4.35  & 3.46 & 3.77 & 0.29 & 3.59 \\
			CsTe  &  6.45 & 4.58  & 3.72 & 3.97 & 0.31 & -1.41 \\
			\hline
			CsP    &  5.79 & 4.22 & 3.34 & 3.65 & 0.28 & -0.35\\
			CsAs & 5.99 & 4.31 & 3.46 & 3.73  & 0.31 & -9.07\\
			\hline
			\hline
		\end{tabular}
	\end{table}

	\begin{table}
		\centering
		\caption{The parameters of the lattice model obtained by fitting the first-principles band structures.  The unit is in eV.}
		\begin{tabular}{ccc}
			\hline
			\hline
			%		-40.54, 8.96254, -352.461, -17.6337, -21.8191, 16.8675, -9.83529
			\multirow{2}*{On-site energy}
			&$e_1$& -0.070\\
			&$e_2$& 0.022\\
			\hline
			\multirow{4}*{Hopping integral} 
			&$t_{11}$& 0.043\\
			&$t_{22}$&  -0.074\\
			&$t_{21}$& -0.053\\
			&$r_{11}$& 0.008\\
			\hline
			\multirow{1}*{SOC strength}
			&$\xi$   &  -0.069 \\ 
			\hline
			\hline
		\end{tabular}
		\label{tbl:model}
	\end{table}

	\newpage

	\begin{figure}
		\includegraphics[width=0.8\columnwidth]{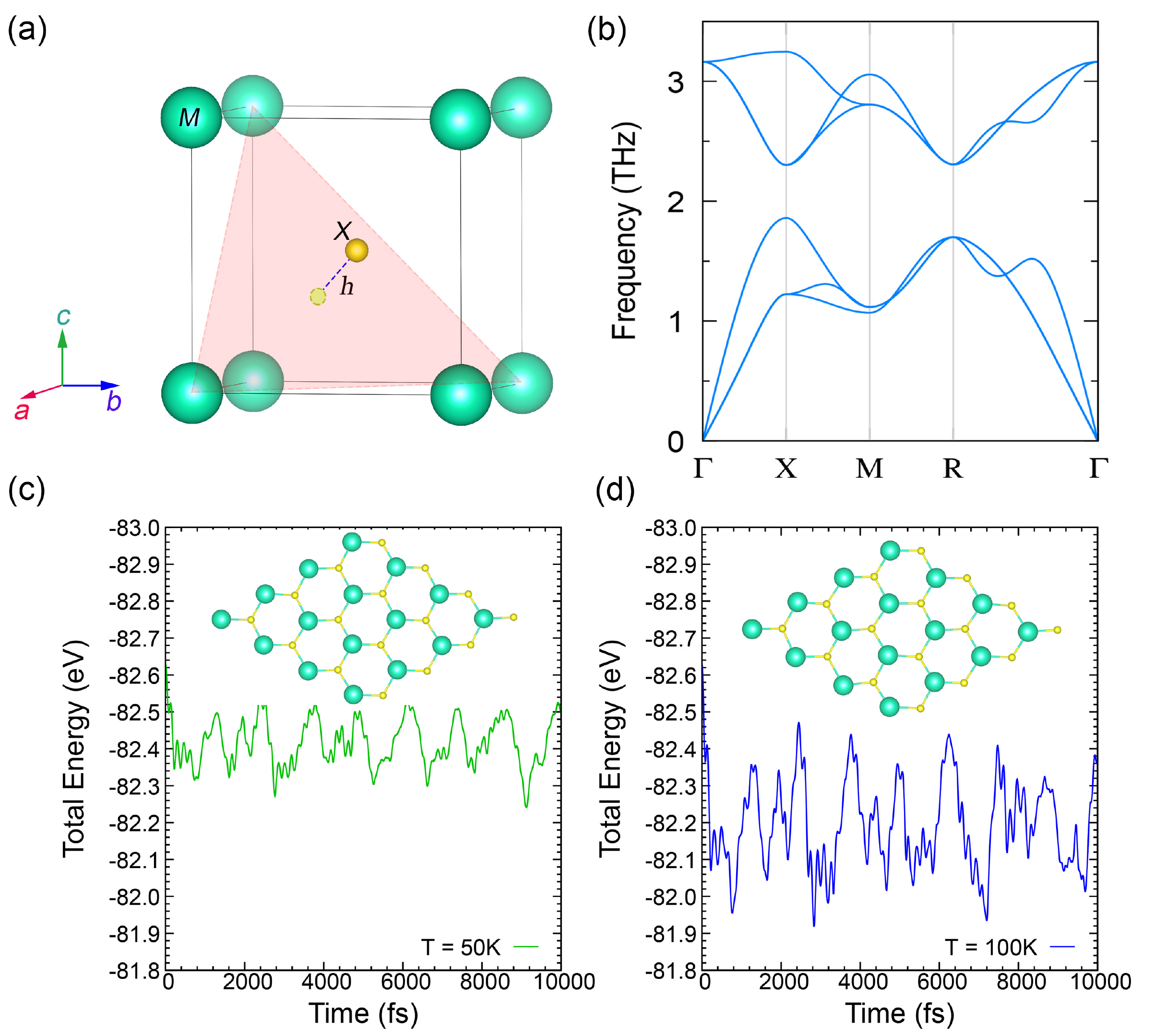}
		\caption{(a) The crystal structure of bulk \textit{MX} (\textit{M} = Li, Na, K, Rb, Cs; \textit{X} = S, Se, Te).  The (111) plane is highlighted by pink color.  The large green and small yellow balls represent \textit{M} and \textit{X} atoms, respectively.  The solid yellow circle represents the projection of \textit{X} atom onto the (111) plane and $h$ is the bulking height.  (b) The phonon spectrum of bulk CsS.  (c,d) The energy evolution of CsS monolayer by employing molecular dynamics simulation at the temperatures of 50 K and 100 K, respectively.  The insets show the snapshots of hexagonal honeycomb lattice.}
	\end{figure} 

	\begin{figure}
		\includegraphics[width=\columnwidth]{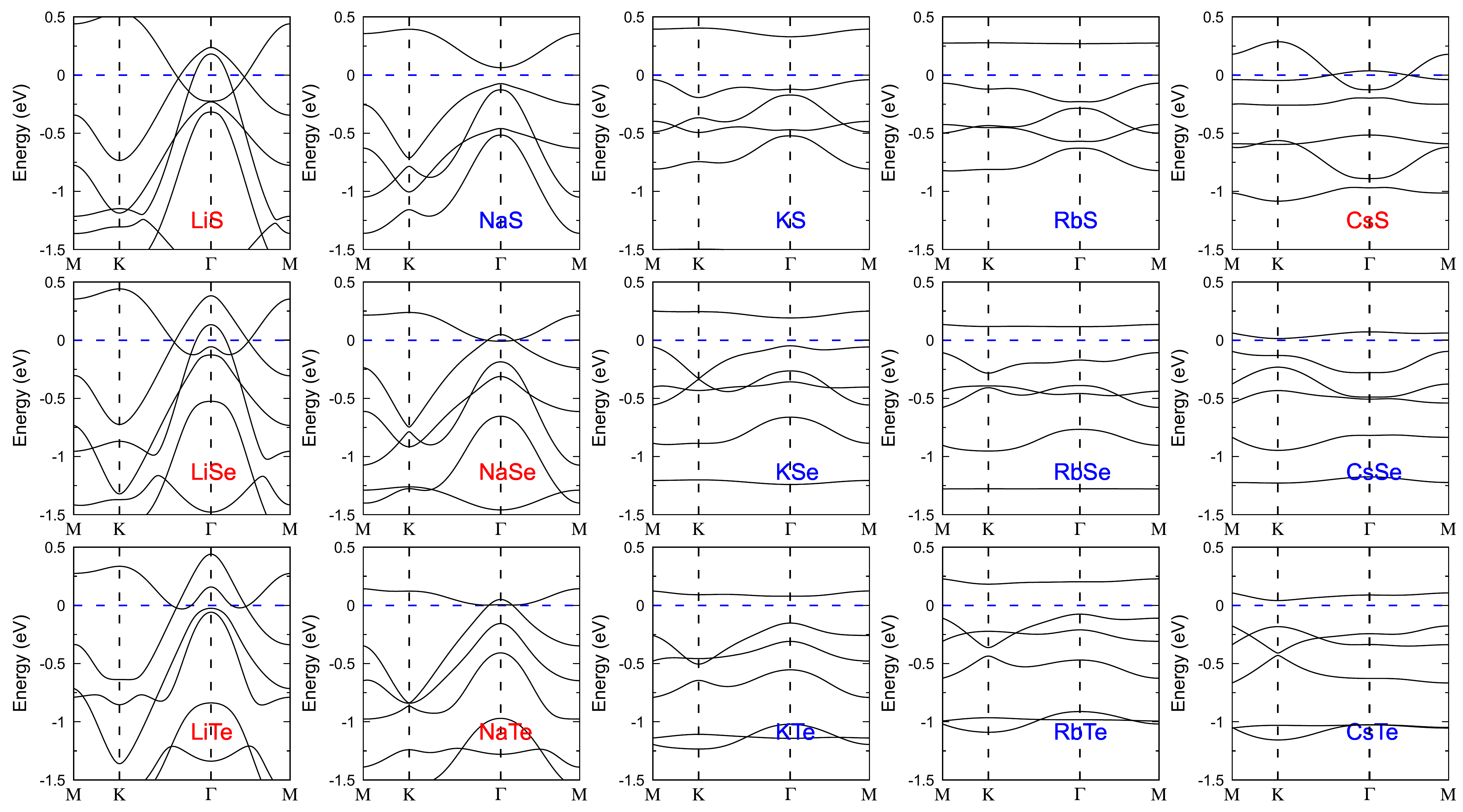}
		\caption{The relativistic band structures of \textit{MX} MLs with the out-of-plane magnetization.  The candidates of nodal loop half-metal are marked in red font.}
	\end{figure}

	\begin{figure}
		\includegraphics[width=0.8\columnwidth]{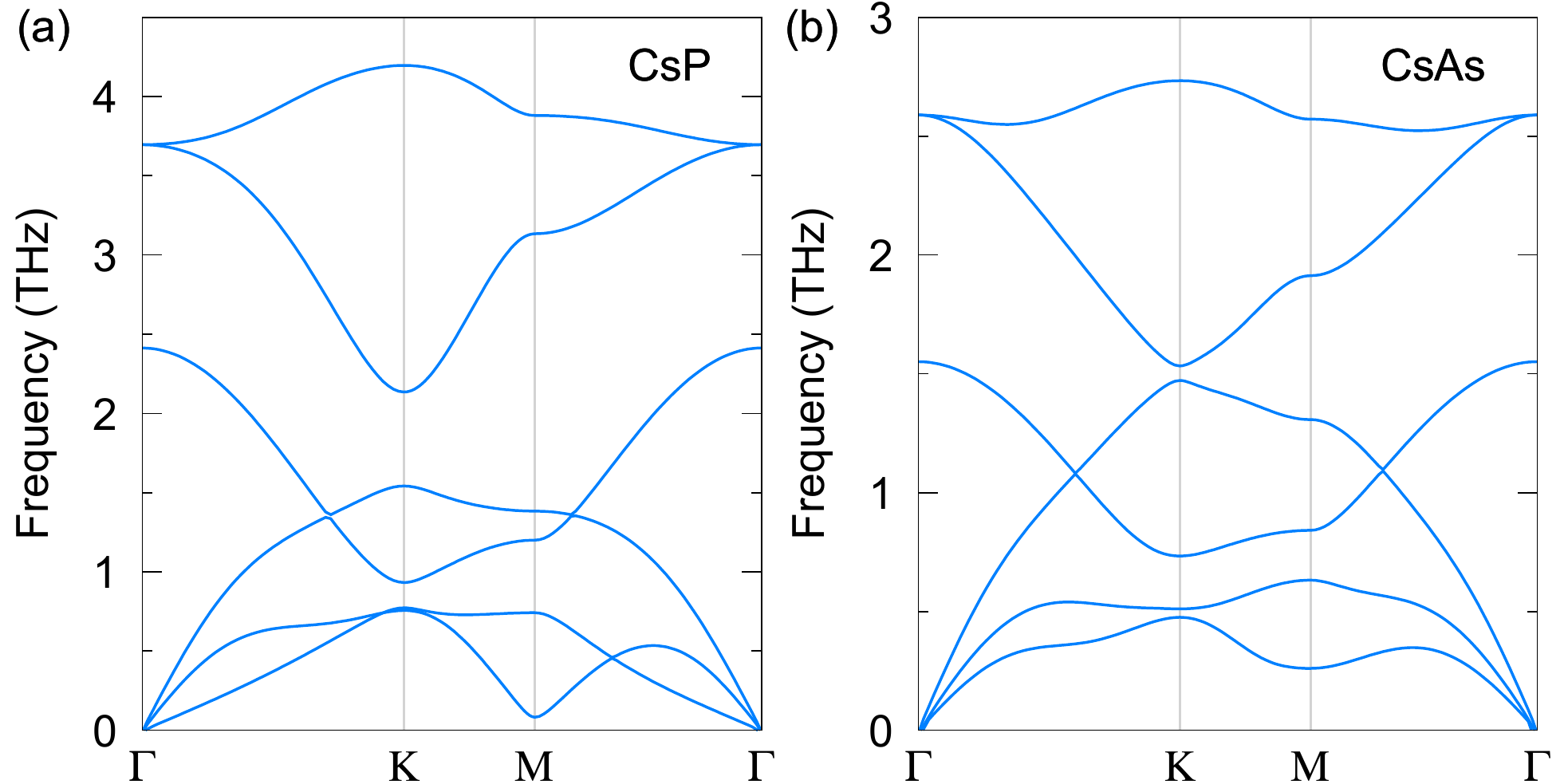}
		\caption{The phonon spectra of CsP and CsAs MLs.}
	\end{figure}

	\clearpage
		
	%\bibliography{refs}
	\providecommand{\latin}[1]{#1}
	\makeatletter
	\providecommand{\doi}
	{\begingroup\let\do\@makeother\dospecials
		\catcode`\{=1 \catcode`\}=2 \doi@aux}
	\providecommand{\doi@aux}[1]{\endgroup\texttt{#1}}
	\makeatother
	\providecommand*\mcitethebibliography{\thebibliography}
	\csname @ifundefined\endcsname{endmcitethebibliography}
	{\let\endmcitethebibliography\endthebibliography}{}

\end{document}